\documentclass[%
reprint,  
superscriptaddress,
showpacs,
preprintnumbers,
showkeys,
 amsmath,amssymb,
 aps,
prb,
]{revtex4-1}
\usepackage{enumitem}
\usepackage{graphicx}
\usepackage{dcolumn}
\usepackage{bm}
\usepackage{hyperref}

\begin{document}


\title{Proton dynamics in high-pressure ice-VII from density functional theory}

\author{Florian Trybel}
\email{f.trybel@uni-bayreuth.de}
\affiliation{ Bayerisches Geoinstitut, Universit{\"a}t Bayreuth, D-95440 Bayreuth, Germany}
\author{Michael Cosacchi}
\affiliation{Theoretische Physik III, Universit{\"a}t Bayreuth, D-95440 Bayreuth, Germany}
 \author{Thomas Meier}
\affiliation{ Bayerisches Geoinstitut, Universit{\"a}t Bayreuth, D-95440 Bayreuth, Germany}
\author{Vollrath Martin Axt}
\affiliation{Theoretische Physik III, Universit{\"a}t Bayreuth, D-95440 Bayreuth, Germany}
\author{Gerd Steinle-Neumann}%
\affiliation{ Bayerisches Geoinstitut, Universit{\"a}t Bayreuth, D-95440 Bayreuth, Germany}

\date{\today}

\begin{abstract} 
Using a density-functional-theory-based approach, we explore the symmetrization and proton dynamics in ice-VII, for which recent high-pressure NMR experiments indicate significant proton dynamics in the pressure-range of $20-95$ GPa. We directly sample the potential seen by the proton and find a continuous transition from double- to single-well character over the pressure range of 2 to 130 GPa accompanied by proton dynamics in agreement with the NMR experiments.  
\end{abstract}

\maketitle
\section{Introduction}

\begin{figure}[h]
\includegraphics[width=0.45\textwidth]{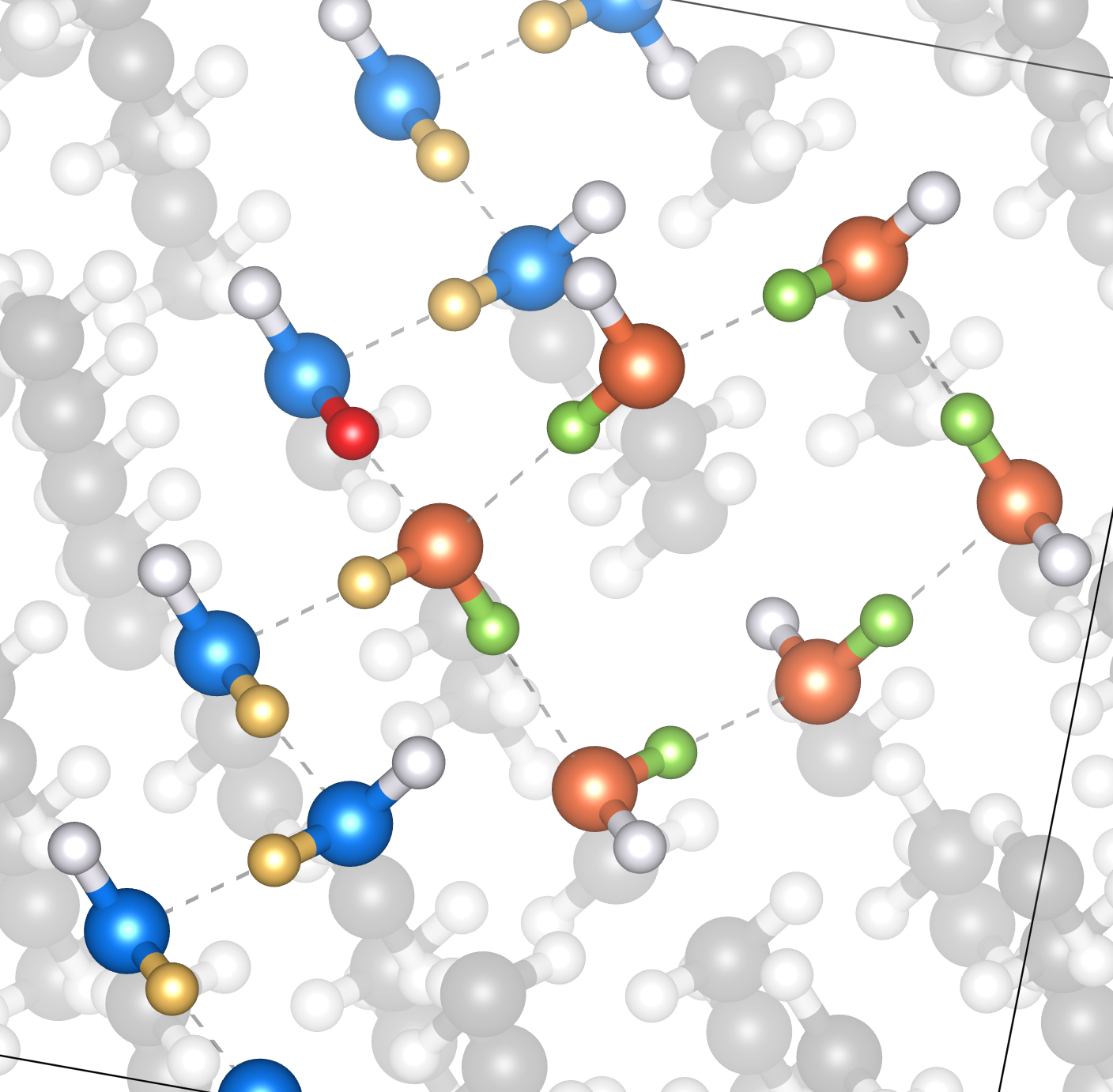}
\caption{Cut through the 4x4x4 ice-VII supercell. The sampling of the correlated hexagonal configuration is performed by moving the green protons in the ring of orange oxygen atoms. For the single proton sampling, the red proton is moved. To provide local charge neutrality, the ice-rule violating defects for the OH$^-$ and H$_3$O$^+$ configurations are moved  to the edge of the simulations cell, along a path similar to that outlined by the golden hydrogen atoms. VESTA 3 \cite{Vesta} is used for structural visualizations.}
\label{fig:sample_vis}
\end{figure}

The  discovery of ice-VII inclusions in diamonds  from the Earth's mantle \cite{Tschauner2018} highlights the importance of this high-pressure phase of water for planetary interiors beyond the icy satellites of Jupiter and Saturn in our solar system \cite{doi:10.1002/2016JE005081,0034-4885-81-6-065901}, and potentially H$_2$O-dominated exosolar planets \cite{Valencia2007,Monteux2018}. At room temperature ($T$), water crystallizes as ice-VII at pressures ($P$) above 2 GPa in a cubic structure (spacegroup Pn$\bar{3}$m), based on a body centered cubic (\textit{bcc}) arrangement of oxygens with two possible proton positions along the diagonal O-O direction (dOOd) that are occupied randomly (Fig. \ref{fig:sample_vis}), but assumed to follow the ice rules: Each oxygen atom is covalently bound to two hydrogen atoms and each resulting water-like unit forms two hydrogen bonds to other oxygen atoms \cite{ice-rules}. With increasing $P$, the O-H$\cdots$O bond continously symmetrizes to form ice-X, a process that has been of great interest in high-$P$ physics and chemistry \cite{Polian1984,Lee1993,Goncharov1996,Aoki1996,Wolanin1997,Bernasconi1998,Benoit1998,Loubeyre1999,Benoit2002, Sanz2004,Caracas2008, Asahara2010, Lu2011, Lin2011,French2015,Tsuchiya2017,Klotz2017,Hernandez2018}. In ice-VII a double-well potential along the O-O direction can be found under correlated proton movement \cite{Benoit1998,Lin2011} which changes significantly under compression. This double-well potential provides the basis for understanding the proton dynamics, recently observed with nuclear magnetic resonance (NMR) experiments in the diamond anvil cell~\cite{Meier2018}. Although numerous studies have investigated the properties of high-$P$ ice-VII for more than 35 years \citep{Polian1984}, no consensus on the symmetrization $P$ has emerged. Proton dynamics and the underlying potential have been calculated from computationally expensive path-integral based simulations at individual $P$ conditions only \cite{Lin2011} (\citet{DrechselgrauMarx2017} for hexagonal ice), or indirectly from density-functional-theory (DFT) based approaches \cite{Benoit2002,Caracas2008,Lu2011, Hernandez2018}.

Using Kohn-Sham (KS) DFT, we trace the potential by displacing all, six and a single proton along dOOd to investigate proton dynamics and explore symmetrization under compression.
\section{DFT-Simulations}
All KS-DFT-based calculations are performed with {\sc Quantum Espresso} 6.1 ~\citep{QE-2009,QE-2017}, where possible using GPU-acceleration \cite{romero2017performance}. We combine the optBK88-vdW~\cite{Becke1988,Dion2004, Thonhauser2007,Langreth2009, Klimes2010, Thonhauser2015, Berland2015} approximation to exchange and correlation, found most suitable for various water properties \citep{gillan2016perspective}, with projector augmented wave atomic files for H and O based on the Perdew-Burke-Ernzerhof \cite{PBE1996} exchange-correlation approximation (1s electrons treated as semi-relativistic core states for O). Convergence tests with a threshold of $10^{-4}$ Ry lead to a Monkhorst-Pack k-point grid \cite{MonkhorstPack} of  1$\times$1$\times$1 for the 384 atom cells and a cutoff energy for plane wave expansion of 130 Ry. 

\begin{figure}[b!]
\begin{center}
\includegraphics[width=0.47\textwidth]{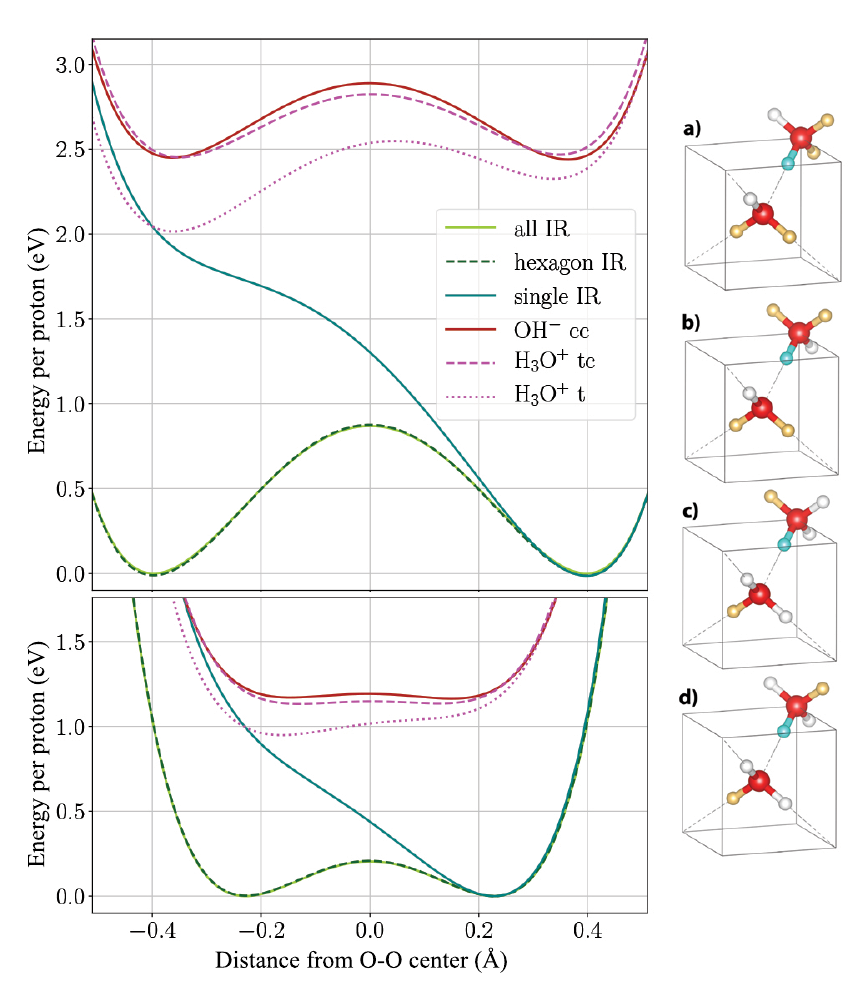}
\caption{Energy of different proton configurations for $l=3.35$ \AA~\textbf{(top)} and $l=2.95$ \AA~\textbf{(bottom)} and \textbf{a)}-\textbf{d)} visualization of possible configurations with a single-proton ice-rule (IR) violation: Oxygen is shown in red, unoccupied hydrogen positions in white, occupied hydrogen positions in yellow and sampled hydrogen in blue: \textbf{a)} H$_3$O$^+$ in cis configuration (cis-H$_3$O$^+$), \textbf{b)} trans-H$_3$O$^+$, \textbf{c)} cis-OH$^-$ and \textbf{d)} trans-OH$^-$. Potentials displayed are for the collective motion of protons (all), the collective hexagonal configuration (hexagon) and a single proton (single), all obeying IR. Potentials for the IR-breaking arrangements are trans-H$_3$O$^+$ (t), without and with conjugation (tc) of the associated charge defect, and cis-OH$^-$ in the conjugated configuration (cc). Conjugated trans-H$_3$O$^+$ and conjugated cis-OH$^-$ are shown as they bracket the potential of all conjugated IR-breaking structures.}
\label{fig:ice-rule breaking}
\end{center}
\end{figure}
\section{Proton Potentials}
We build ice-VII cells with 384 atoms (lattice constants between $l=3.35$ \AA~and $2.55$~\AA), following the ice rules \cite{ice-rules}. All protons are simultaneously moved by the same amount along the respective O-O diagonal (22 steps),  calculated energies are interpolated with a third-order spline, and the optimal proton position at each $l$ is determined. We obtain a double-well potential (characterized by an energy barrier $\Phi_{\rm m}$ and the distance between minimum location and the center of the O-O diagonal, $\delta_{\rm c}$)  at low compression (Fig. \ref{fig:ice-rule breaking}), and we  observe a continuous transition to a single well with decreasing $l$ (Fig. S1 in the Supp. Mat. \cite{Supplement}). In the following, the optimized proton position from the collective displacement of all protons at the respective $l$ is used as the reference position, when only some protons are considered.

Moving a single proton along dOOd in the ice-rule conforming cell (red atom in Fig. \ref{fig:sample_vis}) leads to an asymmetric single-well potential, with the minimum corresponding to one of the minima in the double well obtained from the collective displacement of all protons (Fig. \ref{fig:ice-rule breaking}), which -- once the double-well character is lost -- converges to a symmetric potential (Fig. S2 in the Supp. Mat. \cite{Supplement}).

\begin{figure}[t!]
\includegraphics[width=0.45\textwidth]{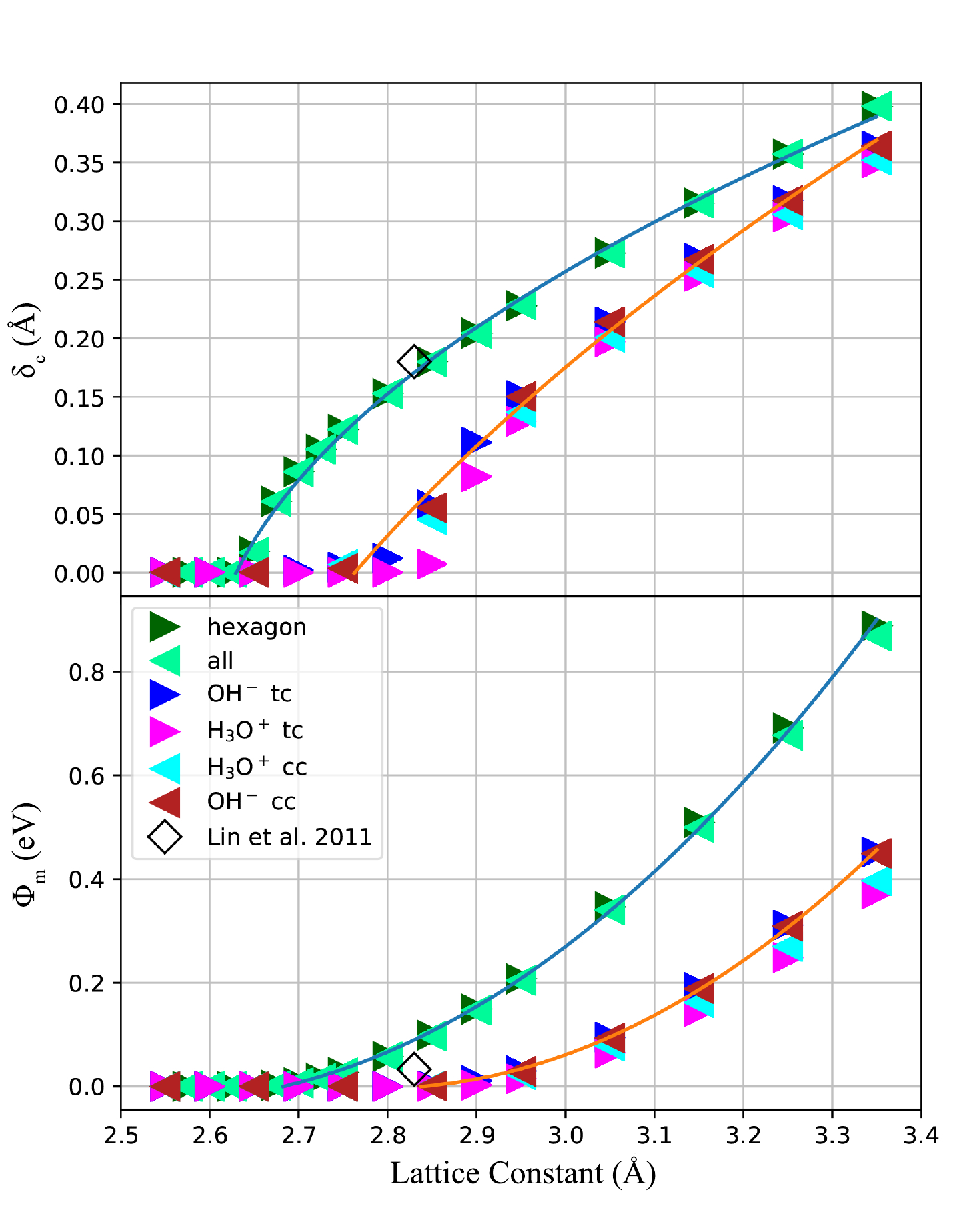}
\caption{Parameters describing the structure of the double-well potential as a function of lattice parameter $l$. \textbf{(top)} $\delta_{\rm c}$, the distance between the center of the double well and the minima. \textbf{(bottom)} $\Phi_{ \rm m}$, the height of the barrier. Solid lines are fits of a square root for $\delta_{\rm c}$ and a quadratic function for $\Phi_{ \rm m}$ to the hexagonal (green, $\Phi^{\rm Hex}_{ \rm m}=1.42(x-2.54)^{2}$, $\delta^{\rm Hex}_{\rm c}=0.57(x-2.59)^{0.5}-0.10$) and trans-OH$^{-}$ points (orange, $\Phi^{ {\rm OH}^-{tc}}_{ \rm m}=1.46(x-2.79)^2$, $\delta^{ {\rm OH}^-{tc}}_{\rm c}=0.90(x-2.49)^{0.5}-0.46$. Open black diamonds show result of path integral \textit{ab-initio} molecular dynamics simulations at $l=2.84$ \AA~\citep{Lin2011}.}
\label{fig:MinMax}
\end{figure}

In previous path-integral-based studies \cite{Lin2011}, a collective motion of six protons in a hexagonal configuration was identified without breaking the ice rules. We follow this suggestion and move six protons in such a configuration (green atoms in Fig. \ref{fig:sample_vis}), while the others remain at the optimized position for the respective $l$. We obtain potentials almost perfectly corresponding to the all-proton case in terms of energy/proton (Fig. \ref{fig:ice-rule breaking}), with the double-well character steadily decreasing with decreasing $l$ (Fig. \ref{fig:MinMax}). Under further compression, the high-$P$ single-well becomes increasingly localized (Fig. S3 in the Supp. Mat. \cite{Supplement}). The potentials show a slight asymmetry caused by the non-central position of the hexagon in the simulation cell in combination with periodic boundary conditions.

By breaking the ice rules, it is possible to create environments in which a single proton experiences a double-well potential when moved along the O-O diagonal.  Four different cases of ice-rule breaking configurations (Fig. \ref{fig:ice-rule breaking}) can be distinguished based on the oxygen-proton ratio, OH$^-$ (Fig. \ref{fig:ice-rule breaking} c,d)  and H$_3$O$^+$ (Fig. \ref{fig:ice-rule breaking} a,b), and different configuration, i.e.,  (rotational) symmetry: (pseudo-)cis  with protons aligned along dOOd  (Fig. \ref{fig:ice-rule breaking} a,c), and trans with protons facing each other  (Fig. \ref{fig:ice-rule breaking} b,d). Due to violation of local charge neutrality, the potentials of all OH$^-$  and H$_3$O$^+$ configurations are higher in energy and asymmetric, as illustrated explicitly for trans-H$_3$O$^+$ in Fig. \ref{fig:ice-rule breaking}. To mitigate the asymmetry, we conjugate the charge defect along a path of water molecules (golden atoms in Fig. \ref{fig:sample_vis}) to the edge of the simulation cell to minimize its influence (Figs. S4 and S5 in the Supp. Mat. \cite{Supplement}). The violation of the ice rules \cite{ice-rules} introduces an energy difference of $\simeq$ 2 eV at $l=3.35$ \AA~(Fig. \ref{fig:ice-rule breaking}), which decreases with compression. Restoring local charge neutrality at the sampling point as described earlier, leads to a small increase of the shift by $\simeq 0.05$ eV (Fig. \ref{fig:ice-rule breaking}).

Comparing $\delta_{\rm c}$ and $\Phi_{\rm m}$ for all six cases discussed above (Fig. \ref{fig:MinMax}), we find a splitting into two groups: Ice-rule violating structures loose the double-well character at a smaller compression, $l \lesssim 2.8-2.85$  \AA, compared to the ice-rule conforming structures, with $l \lesssim 2.675$ \AA. 

Nevertheless, the functional behavior of both parameters with compression, expressed in terms of lattice parameter $l$, is similar for all cases - following a quadratic function for $\Phi_{\rm m}$ and a square root for $\delta_{\rm c}$ (Fig. \ref{fig:MinMax}). Comparing the double-well potential from the path-integral simulations by \citet{Lin2011} with our results, their $\delta_{\rm c}$ is in very good agreement with the hexagonal configuration; $\Phi_{\rm m}$ is larger in our calculations, with the difference most likely caused by theirs being ensemble calculations in contrast to our individual configuration approach and -- more importantly -- the classical treatment of protons in our calculations to this point.

Temperature has been neglected in the evalution of the potentials. As we create unstable configurations during sampling, harmonic lattice dynamics simulations would show negative phonon modes, and in molecular dynamics the structure would fall back to an equilibrium. The good agreement between the path integral results \cite{Lin2011} and our calculations with respect to $\delta_{\rm c}$ suggests that vibrational effects do not influence the width of the potential significantly. By evaluating results as a function of $l$ and not $P$, thermal expansion does not influence the outcome. In order to compare to experimental data, we use the thermal equation of state by \citet{French2015} to transform $l$ to $P$ at 300 K.

\section{Proton Dynamics}

Many high-$P$ techniques do not provide access to the potential itself, but the electron density close to the protons by X-rays \cite{Wolanin1997,Loubeyre1999,Somayazulu2008,Sugimura2008}, and the response of the lattice to an excitation of phonon modes by Raman/infrared \cite{Hirsch1986,Goncharov1996,Wolanin1997,Pruzan2003,Zha2016} or Brillouin \cite{Asahara2010} spectroscopy can be measured at high $P$. 
High-$P$ NMR spectroskopy, by contrast, enables the investigation of effective proton dynamics \cite{Meier2018} by performing a line-shape analysis. 

\subsection{Ansatz}

In order to compare with the latter results, we use a matrix exponential formalism to solve the time ($t$) dependent Schr{\"o}dinger equation for a wave packet in the respective one-dimensional potentials obtained from the DFT-based sampling (see Supp.~Mat. \cite{Supplement}~for details). The initial state is described by a generic 1D-Gaussian of the form 

\begin{align}
\Psi~=~& \mathcal{I}^{-1} \cdot \mathcal{F}(\dfrac{x-x_{\rm 0}}{3a}) \cdot \notag \\
&\cdot \dfrac{1}{\sqrt{a\sqrt{\pi}}}\exp\bigg(-0.5\bigg(\dfrac{x-x_{\rm 0}}{a}\bigg)^2\bigg),
\label{eq:G}
\end{align}
where $\hbar$ is the reduced Planck constant, $x$ the position along dOOd, $a$ and $x_{\rm 0}$ the width and center of the Gaussian, $m$ the proton mass; $\mathcal{I}$ ensures normalization and $\mathcal{F}$ is a mollifier \cite{friedrichs1944}, improving the localization of $\Psi$, which leads to a significant speed-up and higher stability of the numerics, without affecting any physical features. 

We assume that through spontaneous symmetry breaking the protons are located at one of the minima at $t_{\rm 0}$ and we therefore choose the initial state such that $x_{\rm 0}$ coincides with one of the minima of the potential for each $l$.
\begin{figure}[b!]
\begin{center}
\includegraphics[width=0.48\textwidth]{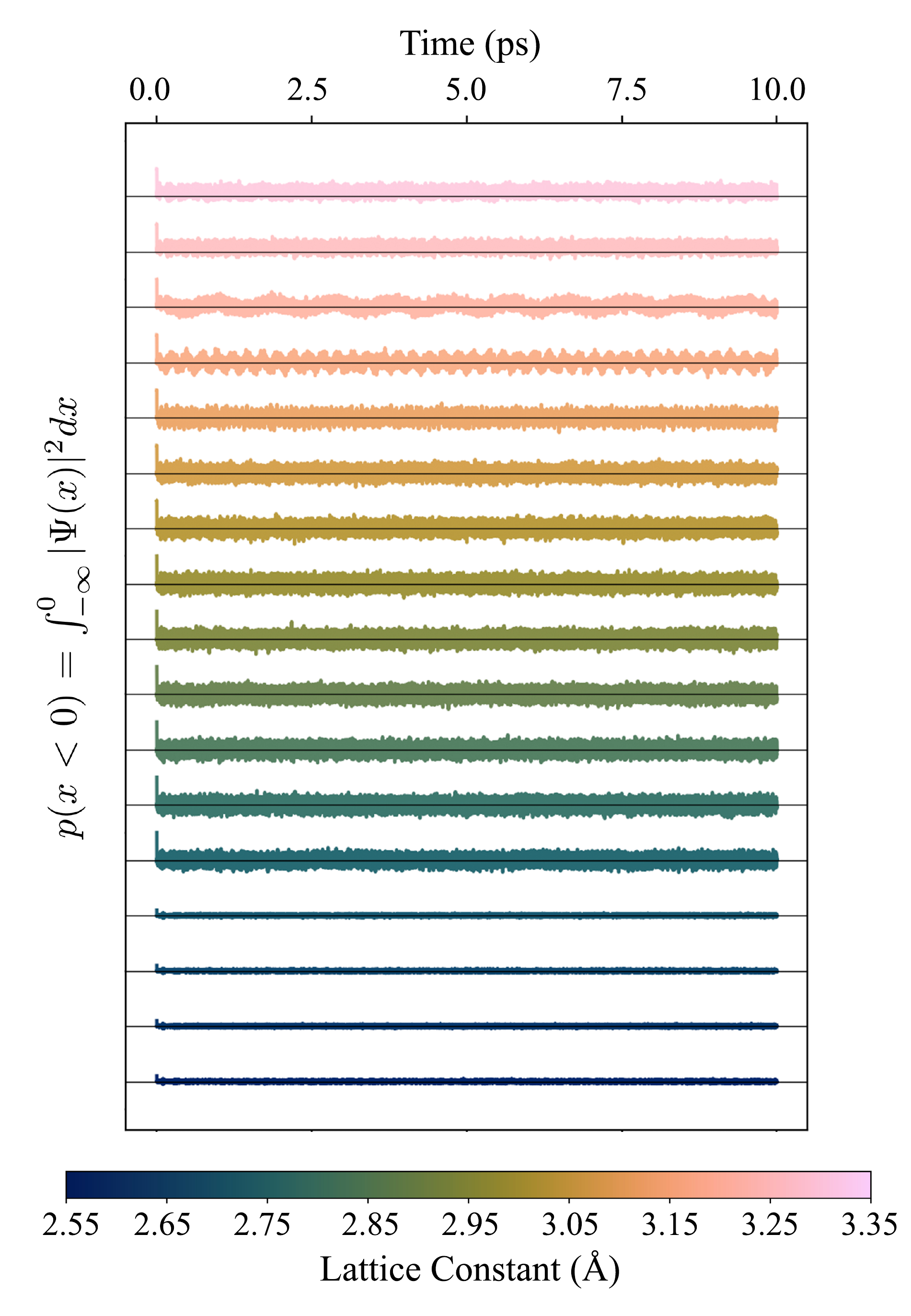}
\caption{ Probability of finding the proton in the left half of the potential ($p(x<0)$) as a function of time for the hexagonal configuration. Black lines indicate $p(x<0)=0.5$. For $l \geq 3.25$ \AA,~ $p(x<0) \gtrsim 0.5$ by construction. With increasing compression (decreasing $l$) the noise level increases, but no clear underlying dynamics is found until at $l \lesssim 3.2$ \AA~ $p(x<0)$ starts to oscillate around a value of 0.5 with decreasing wavelength. At $l \lesssim 2.7$ \AA~ the amplitude of $p(x<0)$ suddenly collapses and no further dynamics is observed.}
\label{fig:dynamics}
\end{center}
\end{figure}
\begin{figure}[t]
\includegraphics[width=0.5\textwidth]{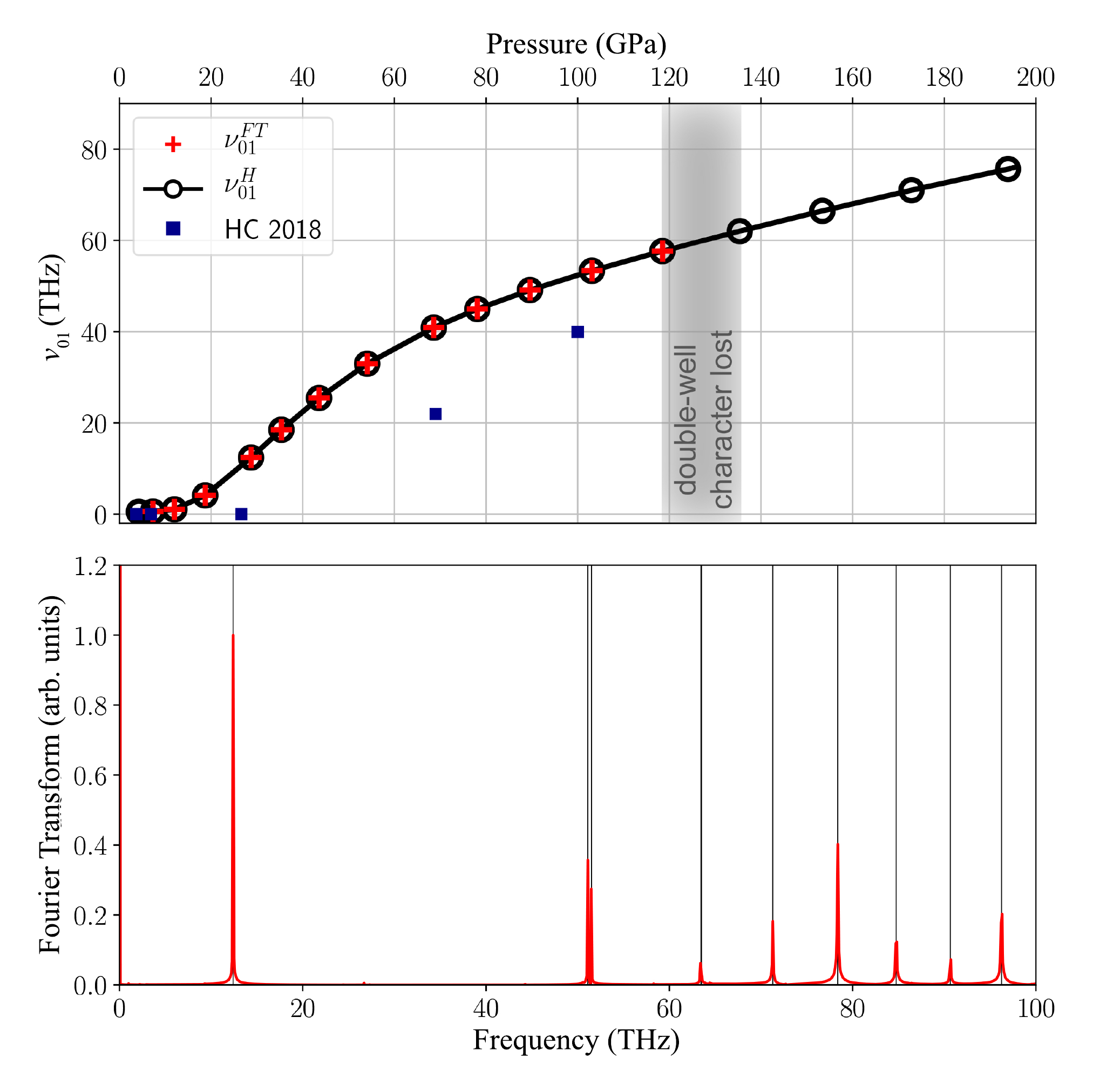}
\caption{ Results of the dynamical calculation for the hexagonal configuration. \textbf{(bottom)} Transition frequency spectrum calculated from the eigenvalues of the Hamiltonian (black vertical lines) and the solution to the dynamic problem, represented by the Fourier transformation of the probability of finding the proton in the left side of the potential, $p(k)$, with amplitudes normalized to $\nu_{01}$ (red) for $l=2.95$ \AA. \textbf{(top)} $\nu_{01}^{H}$ from the eigenvalue analysis (black circles) and position of $\nu_{01}^{FT}$ in $p(k)$ (red crosses) as a function of $P$. Blue squares show hydrogen jump rates calculated by \citet{Hernandez2018} (HC 2018) on the basis of molecular dynamics simulations.}
\label{fig:Tunnel_T}
\end{figure}

An eigenvalue analysis of the Hamiltonian is helpful to understand the idea behind this choice with respect to the analysis of effective single-particle dynamics and the spectral features. If we construct the initial state as a superposition of the two lowest eigenstates, it is straightforward to show that the resulting wave packet oscillates between the minima with a frequency  exactly corresponding to the difference in energies of both eigenstates ($\nu_{01}^{H}$), representing tunnelling.  As the eigenstates of the Hamiltonian are a basis set, any initial state can be decomposed into them, and the resulting dynamics includes all frequencies ($\nu^{H}$) which correspond to the combinatorically possible differences between their energies. While exactly diagonalizing the Hamiltonian yields all $\nu^{H}$ taking part in the dynamics, solving the problem with a physically motivated initial state additionally provides access to amplitudes.

We calculate the $t$-evolution of $\Psi$ with respect to the potentials of the hexagonal as well as the four ice-rule violating configurations \cite{footnote3}. In each $t$-step, the probability of finding the proton in the left half of the double-well potential is given by  (Fig. \ref{fig:dynamics})
\begin{equation}
p(x<0) =\int_{\rm -\infty}^{0} |\Psi(x)|^2 dx.
\end{equation}
The Fourier transform of $p(x<0)$, $p(k)$, results in a $\nu^{FT}$-spectrum that perfectly coincides with the transitions based on the Hamiltonian (Fig. \ref{fig:Tunnel_T} for the hexagonal configuration), which can be seen as a check for the solver, in addition to energy and norm conservation.

Choosing $x_{\rm 0}$ for the initial state as the minimum of the respective potential leads to a non-zero overlap between the ground state and the first excited state and therefore $\nu_{01}^{FT} > 0$ as long as $x_{\rm 0}$ is not central. Once the minimum of the potential, and consequently $x_{\rm 0}$, is at the center of dOOd, no overlap between the ground state (symmetric) and the first excited state (antisymmetric) is expected, and the amplitude of $\nu_{01}^{FT}$ should vanish, indicating the completion of symmetrization.

\subsection{Pressure dependence}
The frequency $\nu_{\rm 01}^{H}$, calculated from the eigenvalues, increases continuously with $P$ in a square root-like fashion between 5 and $\sim 100$-$120$ GPa and linearly upon further $P$-increase to our maximum $P \approx 200$  GPa (Fig. \ref{fig:Tunnel_T} for the hexagonal configuration), indicating that there is no structural phase transition of first or second order. This change in slope indicates a change in dynamic behavior independent of the particular form of the initial state, providing an additional (although less sensitive) analysis tool, and shows that the exact form of $\Psi$, in particular the use of $\mathcal{F}$, only improves the analysis, but is not required. 

\begin{figure}[!h]
\includegraphics[width=0.5\textwidth]{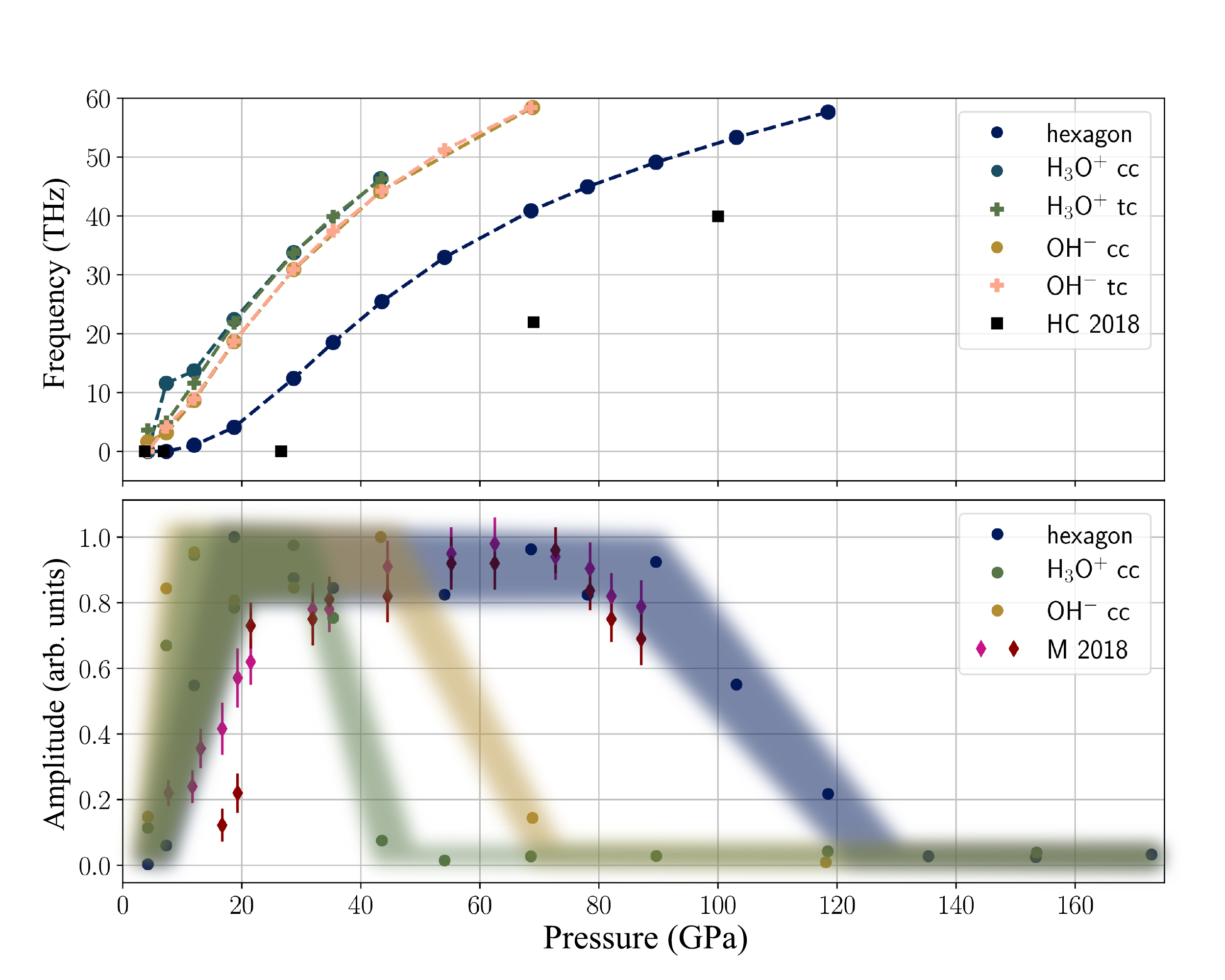}
\caption{
\textbf{(top)} Excitation frequencies $\nu_{01}^{FT}$ as a function of $P$ at 300 K for the different proton configurations explored in this study. Black squares show hydrogen jump rates calculated by \citet{Hernandez2018} (HC 2018) on the basis of molecular dynamics simulations. \textbf{(bottom}) Corresponding amplitudes of $\nu_{01}^{FT}$ in $p(k)$. Amplitudes are normalized to the maximum value for the respective configuration \cite{footnote2}. Diamonds in magenta and purple show the result of a peak analysis of NMR experiments by \citet{Meier2018} (M 2018).}
\label{fig:Tunnel_conf}
\end{figure}

$\nu_{\rm 01}^{FT}$, calculated from $p(k)$, follows this trend to $P \lesssim 130$ GPa, where we find a significant drop in amplitude (Fig. \ref{fig:Tunnel_conf}), suggesting that the $\nu_{\rm 01}$ oscillation is no longer contributing to the dynamics of the system.

Proton jump frequencies calculated by \citet{Hernandez2018} are in quantitative agreement with our results for the hexagonal configuration, but shifted to slightly higher $P$, which is most likely a consequence of their classical treatment of the problem. The occupation of $\nu_{01}^{FT}$ for the ice-rule violating configuration drops at a lower $P$ (Fig. \ref{fig:Tunnel_conf}), reflecting the earlier transition to a single-well potential (Fig. \ref{fig:MinMax}).

\section{Discussion \& Conclusion}
If we compare the amplitude of $\nu_{01}^{FT}$ in $p(k)$ as a function of $P$ with the NMR line shape analysis by \citet{Meier2018} (Fig. \ref{fig:Tunnel_conf}), we find that the hexagonal configuration reproduces the $P$-dependence of the NMR data at room $T$ well: A strong increase in amplitude up to $P \sim 20$ GPa, a plateau to $P\sim 80 - 90$ GPa, and a subsequent drop. This supports the interpretation of \citet{Lin2011} that charge balance tends to be maintained at low $T$, and therefore tunneling is correlated, represented by the hexagonal configuration.

We do not find a sharp phase transition in cubic ice between the emergence of ice-VII at $\sim$ 2 GPa and 200 GPa, well in the stability field of ice-X, but a gradual change from a double- to a single-well potential, and we predict related changes in proton dynamics. Based on our results we can distinguish regions of characteristic proton dynamics and potential (at room $T$): 
\begin{enumerate}[label={(\roman*)}]
\item For $2$ GPa $ < P < 20$ GPa, protons become increasingly delocalized, as the potential barrier decreases;  
\item for $20$ GPa $ < P < 90$ GPa, protons tunnel in a fashion well represented by correlated hexagons;
\item for $90$ GPa $ < P < 130$ GPa, symmetrization occurs and the amplitude of $\nu_{01}^{FT}$ drops significantly as the potential barrier becomes small, but non-zero;
\item for $P \gtrsim$ 130 GPa, the potential is fully symmetric and no proton dynamics is expected for an initially symmetric state. 
\end{enumerate}
In $P$-ranges (i) and (iii), varying energy and momentum transfer to the sample by different experimental methods may modify the probed potential, ranging from an excitation of the proton spin system without significant momentum or energy transfer to the lattice in NMR  \citep{Meier2018}, through momentum transfer by collisions with neutrons \citep{Klotz2017}, the excitation of lattice vibrations in Raman \citep{Hirsch1986,Pruzan2003,Zha2016}, infrared \cite{Goncharov1996} and Brillouin \cite{Asahara2010,Ahart2011,Zhang2019,Li2019} spectroscopy, to high-energy irradiation with X-rays \cite{Wolanin1997,Loubeyre1999,Somayazulu2008,Sugimura2008}.  

The onset of tunneling in $P$-range (i), predicted in our simulations, can be directly linked to low-$P$ features found in X-ray diffraction \cite{Wolanin1997, Loubeyre1999, Somayazulu2008}, but appears to be absent \cite{Goncharov1996, Asahara2010} or only weakly  visible \cite{Li2019,Zha2016} in lattice vibration-based methods. As thermal energy would couple most directly to the phonons, this absence supports the argument that potential parameters and therefore the onset of tunneling does not strongly depend on $T$. Instead, optical methods \cite{Asahara2010, Goncharov1996, Hirsch1986, Zha2016} find anomalies at $P\approx40$ GPa which suggests that momentum transfer interacts with the lattice in a fashion not visible in the ground-state calculations at the basis of our potentials. Features at $P \gtrsim60$ GPa, found in X-ray diffraction \cite{Wolanin1997,Loubeyre1999,Somayazulu2008,Sugimura2008} and optical measurements \cite{Goncharov1996,Asahara2010,Zha2016} correlate with the highest mobility in the NMR data \cite{Meier2018} (Fig. 4), indicating that compressional energy added to the system can no longer be transferred to the protons. 

Optical \cite{Asahara2010, Hirsch1986} and X-ray diffraction studies \cite{Sugimura2008} are in line with our interpretation of $P$-range (iii), with structural changes at $P\gtrsim80$ GPa and $P\approx110$ GPa, respectively. The lower symmetrization $P$ in the optical experiments \cite{Asahara2010, Hirsch1986} supports the argument that momentum transfer may suppress correlated tunneling. The symmetrization $P$ from the X-ray diffraction experiments \cite{Sugimura2008} is in good agreement with our prediction, and the cessation of dynamics extrapolated from the NMR results \cite{Meier2018}. 

Based on this observation we suggest that experimental methods can be expected not to agree on a single transition $P$ in a system like H$_2$O ice. Further, the limited experimental resolution in terms of $P$-sampling may lead to an over-interpretation of the nature of the phase transition, in some cases claming sharp transitions \cite{Asahara2010,Somayazulu2008, Sugimura2008}.
\section*{Acknowledgements}
FT and GSN were supported by Deutsche Forschungsgemeinschaft (DFG) within FOR 2440 (Matter under Planetary Interior Conditions) with grant STE1105/13-1 and TM with grant ME5206/3-1. We thank  F. Ungar (TP III, Universit{\"a}t Bayreuth) for very helpful discussions. Computations were partly performed at the Leibniz Supercomputing Centre of the Bavarian Academy of Sciences and the Humanities. GPU accelerated computations are supported by the NVIDIA Corporation with the donation of a Titan Xp GPU.


\begin{thebibliography}{55}
\bibitem [{foo()}]{footnote3}%
  \BibitemOpen
  \href@noop {} {}\bibinfo {note} {{The potential for the collective movement of all protons is not evaluated as the combined energy barrier of all protons in the system is significantly higher than for the hexagonal collective movement}}\BibitemShut {NoStop}%
\bibitem [{foo()}]{footnote2}%
  \BibitemOpen
  \href@noop {} {}\bibinfo {note} {{For the hexagonal configuration, the point
  at $l=2.85$ \AA~is missing, as the peak in the spectrum is not sufficiently
  resolved by the discrete calculation grid}}\BibitemShut {NoStop}%
\end{thebibliography}
\end{document}